\documentclass[12pt]{iopart}%
\usepackage{iopams}
\usepackage{amsfonts}
\usepackage{amssymb}
\usepackage{graphicx}%

\begin{document}
\title[The Cosmological Constant]{The Cosmological Constant as an Eigenvalue of a Sturm-Liouville Problem and
its Renormalization}
\author{Remo Garattini}

\begin{abstract}
We discuss the case of massive
gravitons and their relation with the cosmological constant, 
considered as an eigenvalue of a Sturm-Liouville problem. A
variational approach with Gaussian trial wave functionals is used as a method
to study such a problem. We approximate the equation to one loop in a
Schwarzschild background and a zeta function regularization is involved to
handle with divergences. The regularization is closely related to the
subtraction procedure appearing in the computation of Casimir energy in a
curved background. A renormalization procedure is introduced to remove the
infinities together with a renormalization group equation. 

\end{abstract}

\address{Universit\`{a} degli Studi di Bergamo, Facolt\`{a} di Ingegneria,\\ Viale
Marconi 5, 24044 Dalmine (Bergamo) ITALY.\\INFN - sezione di Milano, Via Celoria 16, Milan, Italy}
\ead{remo.garattini@unibg.it}

\section{Introduction}

There are two interesting and fundamental questions of Einstein gravity which
have not received an answer yet: one of these is the cosmological constant
$\Lambda_{c}$ and the other one is the existence of gravitons with or without
mass. While the mass-less graviton is a natural consequence of the linearized
Einstein field equations, the massive case is more delicate. At the linearized level,
 we are forced to introduce the Pauli-Fierz mass term\cite{PauliFierz}%
\begin{equation}
S_{P.F.}=\frac{m_{g}^{2}}{8\kappa}\int d^{4}x\sqrt{-g^{\left(  4\right)  }%
}\left[  h^{\mu\nu}h_{\mu\nu}-h^{2}\right]  ,
\end{equation}
where $m_{g}$ is the graviton mass and $\kappa=8\pi G$. $G$ is the Newton
constant. The Pauli-Fierz mass term breaks the symmetry $h_{\mu\nu
}\longrightarrow h_{\mu\nu}+2\nabla_{\left(  \mu\right.  }\xi_{\left.
\nu\right)  }$, but does not introduce ghosts. Boulware and Deser tried to
include a mass in the general framework and not simply in the linearized
theory. They discovered that the theory is unstable and produce
ghosts\cite{BoulwareDeser}. Another problem appearing when one consider a
massive graviton in Minkowski space is the limit $m_{g}\rightarrow0$: the
analytic expression in the massive and in the mass-less limit does not
coincide. This is known as van Dam-Veltman-Zakharov (vDVZ)
discontinuity\cite{vDVZ}. Other than the appearance of a discontinuity in the
mass-less limit, they showed that a comparison with experiment, led the
graviton to be rigorously mass-less. Actually, we know that there exist bounds
on the graviton rest mass that put the upper limit on a value less than
$10^{-62}-10^{-66}g$\cite{mg}. Recently there has been a considerable interest
in massive gravity theories, especially about the vDVZ discontinuity examined
in de Sitter and Anti-de Sitter space. Indeed in a series of papers, it has
been shown that the vDVZ discontinuity disappears in the mass-lees, at least
at the tree level approximation\cite{tree}, while it reappears at one
loop\cite{loop}. If we fix our attention on the positive cosmological term
expanded to one loop, we can see that its structure is%
\begin{equation}
S_{\Lambda_{c}}=\frac{\Lambda_{c}}{4\kappa}\int d^{4}x\sqrt{-g^{\left(
4\right)  }}\left[  h^{\mu\nu}h_{\mu\nu}-\frac{1}{2}h^{2}\right]  ,
\label{lambda_c}%
\end{equation}
which is not of the Pauli-Fierz form\footnote{To this purpose, see also
Ref.\cite{Visser}.}. Nevertheless, we have to note that the non trace terms of
$S_{P.F.}$ and $S_{\Lambda_{c}}$ can be equal if%
\begin{equation}
\frac{m_{g}^{2}}{2}=\Lambda_{c}. \label{mglambda}%
\end{equation}
In other words the graviton mass and the cosmological constant seem to be two
aspects of the same problem. Furthermore, the cosmological
constant suffers the same problem of smallness, because the more recent
estimates on $\Lambda_{c}$ give an order of $10^{-47}GeV^{4}$, while a crude
estimate of the Zero Point Energy (ZPE) of some field of mass $m$ with a
cutoff at the Planck scale gives $E_{ZPE}\approx10^{71}GeV^{4}$ with a
difference of about 118 orders\cite{Lambda}. One interesting way to relate the
cosmological constant to the ZPE is given by the Einstein field equations
withour matter fields%
\begin{equation}
R_{\mu\nu}-\frac{1}{2}g_{\mu\nu}R^{\left(  4\right)  }+\Lambda_{c}g_{\mu\nu
}=G_{\mu\nu}+\Lambda_{c}g_{\mu\nu}=0,
\end{equation}
where $G_{\mu\nu}$ is the Einstein tensor. If we introduce a time-like unit
vector $u^{\mu}$ such that $u\cdot u=-1$, then%
\begin{equation}
G_{\mu\nu}u^{\mu}u^{\mu}=\Lambda_{c}.
\end{equation}
This is simply the Hamiltonian constraint written in terms of equation of
motion. However, we would like to compute not $\Lambda_{c}$, but its
expectation value $\left\langle \Lambda_{c}\right\rangle $ on some trial wave
functional. On the other hand%
\begin{equation}
\frac{\sqrt{g}}{2\kappa}G_{\mu\nu}u^{\mu}u^{\mu}=\frac{\sqrt{g}}{2\kappa
}R+\frac{2\kappa}{\sqrt{g}}\left(  \frac{\pi^{2}}{2}-\pi^{\mu\nu}\pi_{\mu\nu
}\right)  =-\mathcal{H},
\end{equation}
where $R$ is the scalar curvature in three dimensions. Therefore%
\begin{equation}
\frac{\left\langle \Lambda_{c}\right\rangle }{\kappa}=-\frac{1}{V}\left\langle
\int_{\Sigma}d^{3}x\mathcal{H}\right\rangle =-\frac{1}{V}\left\langle
\int_{\Sigma}d^{3}x\hat{\Lambda}_{\Sigma}\right\rangle ,
\end{equation}
where the last expression stands for%
\begin{equation}
\frac{1}{V}\frac{\int\mathcal{D}\left[  g_{ij}\right]  \Psi^{\ast}\left[
g_{ij}\right]  \int_{\Sigma}d^{3}x\mathcal{H}\Psi\left[  g_{ij}\right]  }%
{\int\mathcal{D}\left[  g_{ij}\right]  \Psi^{\ast}\left[  g_{ij}\right]
\Psi\left[  g_{ij}\right]  }=\frac{1}{V}\frac{\left\langle \Psi\left\vert
\int_{\Sigma}d^{3}x\hat{\Lambda}_{\Sigma}\right\vert \Psi\right\rangle
}{\left\langle \Psi|\Psi\right\rangle }=-\frac{\Lambda}{\kappa},
\label{expect}%
\end{equation}
and where we have integrated over the hypersurface $\Sigma$, divided by its volume and
functionally integrated over quantum fluctuation. Note that Eq.$\left(
\ref{expect}\right)  $ can be derived starting with the Wheeler-De Witt
equation (WDW) \cite{De Witt} which represents invariance under \textit{time}
reparametrization. Extracting the TT tensor contribution from Eq.$\left(
\ref{expect}\right)  $ approximated to second order in perturbation of the
spatial part of the metric into a background term, $\bar{g}_{ij}$, and a
perturbation, $h_{ij}$, we get%
\[
\hat{\Lambda}_{\Sigma}^{\bot}=
\]%
\begin{equation}
\frac{1}{4V}\int_{\Sigma}d^{3}x\sqrt{\bar{g}}G^{ijkl}\left[  \left(
2\kappa\right)  K^{-1\bot}\left(  x,x\right)  _{ijkl}+\frac{1}{\left(
2\kappa\right)  }\left(  \triangle_{2}\right)  _{j}^{a}K^{\bot}\left(
x,x\right)  _{iakl}\right].  \label{p22}%
\end{equation}
The propagator $K^{\bot}\left(  x,x\right)  _{iakl}$ can be represented as
\begin{equation}
K^{\bot}\left(  \overrightarrow{x},\overrightarrow{y}\right)  _{iakl}:=%
{\displaystyle\sum_{\tau}}
\frac{h_{ia}^{\left(  \tau\right)  \bot}\left(  \overrightarrow{x}\right)
h_{kl}^{\left(  \tau\right)  \bot}\left(  \overrightarrow{y}\right)
}{2\lambda\left(  \tau\right)  }, \label{proptt}%
\end{equation}
where $h_{ia}^{\left(  \tau\right)  \bot}\left(  \overrightarrow{x}\right)  $
are the eigenfunctions of $\triangle_{2}$. $\tau$ denotes a complete set of
indices and $\lambda\left(  \tau\right)  $ are a set of variational parameters
to be determined by the minimization of Eq.$\left(  \ref{p22}\right)  $. The
expectation value of $\hat{\Lambda}_{\Sigma}^{\bot}$ is easily obtained by
inserting the form of the propagator into Eq.$\left(  \ref{p22}\right)  $ and
minimizing with respect to the variational function $\lambda_{i}\left(
\tau\right)  $. Thus the total one loop energy density for TT tensors is%
\begin{equation}
\Lambda\left(  \lambda_{i}\right)  =-\kappa\frac{1}{4}%
{\displaystyle\sum_{\tau}}
\left[  \sqrt{\omega_{1}^{2}\left(  \tau\right)  }+\sqrt{\omega_{2}^{2}\left(
\tau\right)  }\right]  .
\end{equation}
The above expression makes sense only for $\omega_{i}^{2}\left(  \tau\right)
>0$. To further proceed, we count the number of modes with frequency less than
$\omega_{i}$, $i=1,2$. This is given approximately by%
\begin{equation}
\tilde{g}\left(  \omega_{i}\right)  =\int\nu_{i}\left(  l,\omega_{i}\right)
\left(  2l+1\right)  ,
\end{equation}
where $\nu_{i}\left(  l,\omega_{i}\right)  $, $i=1,2$ is the number of nodes
in the mode with $\left(  l,\omega_{i}\right)  $, such that $\left(  r\equiv
r\left(  x\right)  \right)  $
\begin{equation}
\nu_{i}\left(  l,\omega_{i}\right)  =\frac{1}{2\pi}\int_{-\infty}^{+\infty
}dx\sqrt{k_{i}^{2}\left(  r,l,\omega_{i}\right)  }.
\end{equation}
Here it is understood that the integration with respect to $x$ and $l$ is
taken over those values which satisfy $k_{i}^{2}\left(  r,l,\omega_{i}\right)
\geq0,$ $i=1,2$. Thus the one loop total energy for TT tensors becomes%
\begin{equation}
\frac{1}{8\pi}\sum_{i=1}^{2}\int_{-\infty}^{+\infty}dx\left[  \int
_{0}^{+\infty}\omega_{i}\frac{d\tilde{g}\left(  \omega_{i}\right)  }%
{d\omega_{i}}d\omega_{i}\right]  . \label{1loop}%
\end{equation}

\section{The massive graviton transverse traceless (TT) spin 2 operator for the Schwarzschild
metric and the W.K.B. approximation}

The further step is the evaluation of Eq.$\left(  \ref{1loop}\right)  $, when
the graviton has a rest mass. Following Rubakov\cite{Rubakov}, the Pauli-Fierz
term can be rewritten in such a way to explicitly violate Lorentz symmetry,
but to preserve the three-dimensional Euclidean symmetry. In Minkowski space
it takes the form
\begin{equation}
S_{m}=-\frac{1}{8\kappa}\int_{\mathcal{M}}d^{4}x\sqrt{-g}\mathcal{L}_{m},
\end{equation}
where%
\begin{equation}
\mathcal{L}_{m}=m_{0}^{2}h^{00}h_{00}+2m_{1}^{2}h^{0i}h_{0i}-m_{2}^{2}%
h^{ij}h_{ij}+m_{3}^{2}h^{ii}h_{jj}-2m_{4}^{2}h^{00}h_{ii}%
\end{equation}
A comparison between $S_{m}$ and the Pauli-Fierz term shows that they can be
set equal if we make the following choice\footnote{See also
Dubovski\cite{Dubovsky} for a detailed discussion about the different choices
of $m_{1}$, $m_{2}$, $m_{3}$ and $m_{4}$}%
\begin{equation}
m_{0}^{2}=0\hspace{2cm}m_{1}^{2}=m_{2}^{2}=m_{3}^{2}=m_{4}^{2}=m^{2}>0.
\end{equation}
If we fix the attention on the very special case $m_{0}^{2}=m_{1}^{2}%
=m_{3}^{2}=m_{4}^{2}=0$; $m_{2}^{2}=m^{2}>0$, we can see that the trace part
disappears and we get
\begin{equation}
S_{m}=\frac{m_{g}^{2}}{8\kappa}\int d^{4}x\sqrt{-\hat{g}}\left[  h^{ij}%
h_{ij}\right]  \qquad\Longrightarrow\qquad\mathcal{H}_{m}=-\frac{m_{g}^{2}%
}{8\kappa}\int d^{3}xN\sqrt{\hat{g}}\left[  h^{ij}h_{ij}\right]  .
\end{equation}
Its contribution to the Spin-two operator for the Schwarzschild metric will be%
\begin{equation}
\left(  \triangle_{2}h^{TT}\right)  _{i}^{j}:=-\left(  \triangle_{T}%
h^{TT}\right)  _{i}^{j}+2\left(  Rh^{TT}\right)  _{i}^{j}+\left(  m_{g}%
^{2}h^{TT}\right)  _{i}^{j}\label{spin2}%
\end{equation}
and%
\begin{equation}
-\left(  \triangle_{T}h^{TT}\right)  _{i}^{j}=-\triangle_{S}\left(
h^{TT}\right)  _{i}^{j}+\frac{6}{r^{2}}\left(  1-\frac{2MG}{r}\right)  \left(
h^{TT}\right)  _{i}^{j}.\label{tlap}%
\end{equation}
$\triangle_{S}$ is the scalar curved Laplacian, whose form is%
\begin{equation}
\triangle_{S}=\left(  1-\frac{2MG}{r}\right)  \frac{d^{2}}{dr^{2}}+\left(
\frac{2r-3MG}{r^{2}}\right)  \frac{d}{dr}-\frac{L^{2}}{r^{2}}\label{slap}%
\end{equation}
and $R_{j}^{a}$ is the mixed Ricci tensor whose components are:
\begin{equation}
R_{i}^{a}=\left\{  -\frac{2MG}{r^{3}},\frac{MG}{r^{3}},\frac{MG}{r^{3}%
}\right\}  .
\end{equation}
This implies that the scalar curvature is traceless. We are therefore led to
study the following eigenvalue equation
\begin{equation}
\left(  \triangle_{2}h^{TT}\right)  _{i}^{j}=\omega^{2}h_{j}^{i}\label{p31}%
\end{equation}
where $\omega^{2}$ is the eigenvalue of the corresponding equation. In doing
so, we follow Regge and Wheeler in analyzing the equation as modes of definite
frequency, angular momentum and parity\cite{Regge Wheeler}. In particular, our
choice for the three-dimensional gravitational perturbation is represented by
its even-parity form%
\begin{equation}
\left(  h^{even}\right)  _{j}^{i}\left(  r,\vartheta,\phi\right)  =diag\left[
H\left(  r\right)  ,K\left(  r\right)  ,L\left(  r\right)  \right]
Y_{lm}\left(  \vartheta,\phi\right)  .\label{pert}%
\end{equation}
Defining reduced fields and passing to the proper geodesic distance from the
\textit{throat} of the bridge, the system $\left(  \ref{p31}\right)  $ becomes%

\begin{equation}
\left\{
\begin{array}
[c]{c}%
\left[  -\frac{d^{2}}{dx^{2}}+\frac{l\left(  l+1\right)  }{r^{2}}+m_{1}%
^{2}\left(  r\right)  \right]  f_{1}\left(  x\right)  =\omega_{1,l}^{2}%
f_{1}\left(  x\right) \\
\\
\left[  -\frac{d^{2}}{dx^{2}}+\frac{l\left(  l+1\right)  }{r^{2}}+m_{2}%
^{2}\left(  r\right)  \right]  f_{2}\left(  x\right)  =\omega_{2,l}^{2}%
f_{2}\left(  x\right)
\end{array}
\right.  \label{p34}%
\end{equation}
where we have defined $r\equiv r\left(  x\right)  $ and%
\begin{equation}
\left\{
\begin{array}
[c]{c}%
m_{1}^{2}\left(  r\right)  =m_{g}^{2}+U_{1}\left(  r\right)  =m_{g}^{2}%
+m_{1}^{2}\left(  r,M\right)  -m_{2}^{2}\left(  r,M\right) \\
\\
m_{2}^{2}\left(  r\right)  =m_{g}^{2}+U_{2}\left(  r\right)  =m_{g}^{2}%
+m_{1}^{2}\left(  r,M\right)  +m_{2}^{2}\left(  r,M\right)
\end{array}
\right.  .
\end{equation}
$m_{1}^{2}\left(  r,M\right)  \rightarrow0$ when $r\rightarrow\infty$ or
$r\rightarrow2MG$ and $m_{2}^{2}\left(  r,M\right)  =3MG/r^{3}$. Note that,
while $m_{2}^{2}\left(  r\right)  $ is constant in sign, $m_{1}^{2}\left(
r\right)  $ is not. Indeed, for the critical value $\bar{r}=5MG/2$, $m_{1}%
^{2}\left(  \bar{r}\right)  =m_{g}^{2}$ and in the range $\left(
2MG,5MG/2\right)  $ for some values of $m_{g}^{2}$, $m_{1}^{2}\left(  \bar
{r}\right)  $ can be negative. It is interesting therefore concentrate in this
range, where $m_{1}^{2}\left(  r,M\right)  $ vanishes when compared with
$m_{2}^{2}\left(  r,M\right)  $. So, in a first approximation we can write%
\begin{equation}
\left\{
\begin{array}
[c]{c}%
m_{1}^{2}\left(  r\right)  \simeq m_{g}^{2}-m_{2}^{2}\left(  r_{0},M\right)
=m_{g}^{2}-m_{0}^{2}\left(  M\right) \\
\\
m_{2}^{2}\left(  r\right)  \simeq m_{g}^{2}+m_{2}^{2}\left(  r_{0},M\right)
=m_{g}^{2}+m_{0}^{2}\left(  M\right)
\end{array}
\right.  ,
\end{equation}
where we have defined a parameter $r_{0}>2MG$ and $m_{0}^{2}\left(  M\right)
=3MG/r_{0}^{3}$. The main reason for introducing a new parameter resides in
the fluctuation of the horizon that forbids any kind of approach. It is now
possible to explicitly evaluate Eq.$\left(  \ref{1loop}\right)  $ in terms of
the effective mass. One gets%
\begin{equation}
\Lambda=\rho_{1}+\rho_{2}=-\frac{\kappa}{16\pi^{2}}\sum_{i=1}^{2}\int
_{\sqrt{m_{i}^{2}\left(  r\right)  }}^{+\infty}\omega_{i}^{2}\sqrt{\omega
_{i}^{2}-m_{i}^{2}\left(  r\right)  }d\omega_{i}, \label{tote1loop}%
\end{equation}
where we have included an additional $4\pi$ coming from the angular integration.

\section{One loop energy Regularization and Renormalization}

Here, we use the zeta function regularization method to compute the energy
densities $\rho_{1}$ and $\rho_{2}$. Note that this procedure is completely
equivalent to the subtraction procedure of the Casimir energy computation
where the zero point energy (ZPE) in different backgrounds with the same
asymptotic properties is involved. To this purpose, we introduce the
additional mass parameter $\mu$ in order to restore the correct dimension for
the regularized quantities. Such an arbitrary mass scale emerges unavoidably
in any regularization scheme. Then we have%
\begin{equation}
\rho_{i}\left(  \varepsilon\right)  =\frac{1}{16\pi^{2}}\mu^{2\varepsilon}%
\int_{\sqrt{m_{i}^{2}\left(  r\right)  }}^{+\infty}d\omega_{i}\frac{\omega
_{i}^{2}}{\left(  \omega_{i}^{2}-m_{i}^{2}\left(  r\right)  \right)
^{\varepsilon-\frac{1}{2}}}. \label{zeta}%
\end{equation}
The integration has to be meant in the range where $\omega_{i}^{2}-m_{i}%
^{2}\left(  r\right)  \geq0$. One gets%
\begin{equation}
\rho_{i}\left(  \varepsilon\right)  =\kappa\frac{m_{i}^{2}\left(  r\right)
}{256\pi^{2}}\left[  \frac{1}{\varepsilon}+\ln\left(  \frac{\mu^{2}}{m_{i}%
^{2}\left(  r\right)  }\right)  +2\ln2-\frac{1}{2}\right]  , \label{zeta1}%
\end{equation}
$i=1,2$. To handle with the divergent energy density we extract the divergent
part of $\Lambda$, in the limit $\varepsilon\rightarrow0$ and we set%
\begin{equation}
\Lambda^{div}=\frac{G}{32\pi\varepsilon}\left(  m_{1}^{4}\left(  r\right)
+m_{2}^{4}\left(  r\right)  \right)  .
\end{equation}
Thus, the renormalization is performed via the absorption of the divergent
part into the re-definition of the bare classical constant $\Lambda$
\begin{equation}
\Lambda\rightarrow\Lambda_{0}+\Lambda^{div}.
\end{equation}

The remaining finite value for the cosmological constant reads%
\[
\frac{\Lambda_{0}}{8\pi G}=\frac{1}{256\pi^{2}}\left\{  m_{1}^{4}\left(
r\right)  \left[  \ln\left(  \frac{\mu^{2}}{\left\vert m_{1}^{2}\left(
r\right)  \right\vert }\right)  +2\ln2-\frac{1}{2}\right]  \right.
\]

\begin{equation}
\left.  +m_{2}^{4}\left(  r\right)  \left[  \ln\left(  \frac{\mu^{2}}%
{m_{2}^{2}\left(  r\right)  }\right)  +2\ln2-\frac{1}{2}\right]  \right\}
=\left(  \rho_{1}\left(  \mu\right)  +\rho_{2}\left(  \mu\right)  \right)
=\rho_{eff}^{TT}\left(  \mu,r\right)  . \label{lambda0}%
\end{equation}

The quantity in Eq.$\left(  \ref{lambda0}\right)  $ depends on the arbitrary
mass scale $\mu.$ It is appropriate to use the renormalization group equation
to eliminate such a dependence. To this aim, we impose that\cite{RGeq}%

\begin{equation}
\frac{1}{8\pi G}\mu\frac{\partial\Lambda_{0}^{TT}\left(  \mu\right)
}{\partial\mu}=\mu\frac{d}{d\mu}\rho_{eff}^{TT}\left(  \mu,r\right)  .
\label{rg}%
\end{equation}
Solving it we find that the renormalized constant $\Lambda_{0}$ should be
treated as a running one in the sense that it varies provided that the scale
$\mu$ is changing%

\begin{equation}
\Lambda_{0}\left(  \mu,r\right)  =\Lambda_{0}\left(  \mu_{0},r\right)
+\frac{G}{16\pi}\left(  m_{1}^{4}\left(  r\right)  +m_{2}^{4}\left(  r\right)
\right)  \ln\frac{\mu}{\mu_{0}}. \label{lambdamu}%
\end{equation}
Substituting Eq.$\left(  \ref{lambdamu}\right)  $ into Eq.$\left(
\ref{lambda0}\right)  $ we find%
\[
\frac{\Lambda_{0}\left(  \mu_{0},r\right)  }{8\pi G}=-\frac{1}{256\pi^{2}%
}\left\{  \left(  m_{g}^{2}-m_{0}^{2}\left(  M\right)  \right)  ^{2}\left[
\ln\left(  \frac{\left\vert m_{g}^{2}-m_{0}^{2}\left(  M\right)  \right\vert
}{\mu_{0}^{2}}\right)  -2\ln2+\frac{1}{2}\right]  \right.
\]%
\begin{equation}
\left.  +\left(  m_{g}^{2}+m_{0}^{2}\left(  M\right)  \right)  ^{2}\left[
\ln\left(  \frac{m_{g}^{2}+m_{0}^{2}\left(  M\right)  }{\mu_{0}^{2}}\right)
-2\ln2+\frac{1}{2}\right]  \right\}  . \label{lambdamu0a}%
\end{equation}
We can now discuss three cases: $\left.  1\right)  $ $m_{g}^{2}\gg m_{0}%
^{2}\left(  M\right)  $, $\left.  2\right)  $ $m_{g}^{2}=m_{0}^{2}\left(
M\right)  $, $\left.  3\right)  $ $m_{g}^{2}\ll m_{0}^{2}\left(  M\right)
.$ In case 1), we can rearrange Eq.$\left(  \ref{lambdamu0a}\right)  $ to
obtain%
\begin{equation}
\frac{\Lambda_{0}\left(  \mu_{0},r\right)  }{8\pi G}\simeq-\frac{m_{g}^{4}%
}{128\pi^{2}}\left[  \ln\left(  \frac{m_{g}^{2}}{4\mu_{M}^{2}}\right)
+\frac{1}{2}\right]  ,
\end{equation}
where we have introduced an intermediate scale defined by%
\begin{equation}
\mu_{M}^{2}=\mu_{0}^{2}\exp\left(  -\frac{3m_{0}^{4}\left(  M\right)  }%
{2m_{g}^{4}}\right)  . \label{newscale}%
\end{equation}
With the help of Eq.$\left(  \ref{newscale}\right)  $, the computation of the
minimum of $\Lambda_{0}$ is more simple. Indeed, if we define%
\begin{equation}
x=\frac{m_{g}^{2}}{4\mu_{M}^{2}}\qquad\Longrightarrow\qquad\Lambda
_{0,M}\left(  \mu_{0},x\right)  =-\frac{G\mu_{M}^{4}}{\pi}x^{2}\left[
\ln\left(  x\right)  +\frac{1}{2}\right]  . \label{LambdansM}%
\end{equation}
As a function of $x$, $\Lambda_{0,M}\left(  \mu_{0},x\right)  $ vanishes for
$x=0$ and $x=\exp\left(  -\frac{1}{2}\right)  $ and when $x\in\left[
0,\exp\left(  -\frac{1}{2}\right)  \right]  $, $\Lambda_{0,M}\left(  \mu
_{0},x\right)  \geq0$. It has a maximum for
\begin{equation}
\bar{x}=\frac{1}{e}\qquad\Longleftrightarrow\qquad m_{g}^{2}=\frac{4\mu
_{M}^{2}}{e}=\frac{4\mu_{0}^{2}}{e}\exp\left(  -\frac{3m_{0}^{4}\left(
M\right)  }{2m_{g}^{4}}\right)
\end{equation}
and its value is%
\begin{equation}
\Lambda_{0,M}\left(  \mu_{0},\bar{x}\right)  =\frac{G\mu_{M}^{4}}{2\pi e^{2}%
}=\frac{G\mu_{0}^{4}}{2\pi e^{2}}\exp\left(  -\frac{3m_{0}^{4}\left(
M\right)  }{m_{g}^{4}}\right)
\end{equation}
or
\begin{equation}
\Lambda_{0,M}\left(  \mu_{0},\bar{x}\right)  =\frac{G}{32\pi}m_{g}^{4}%
\exp\left(  \frac{3m_{0}^{4}\left(  M\right)  }{m_{g}^{4}}\right)  .
\end{equation}
In case 2), Eq.$\left(  \ref{lambdamu0a}\right)  $ becomes%
\begin{equation}
\frac{\Lambda_{0}\left(  \mu_{0},r\right)  }{8\pi G}\simeq\frac{\Lambda
_{0}\left(  \mu_{0}\right)  }{8\pi G}=-\frac{m_{g}^{4}}{128\pi^{2}}\left[
\ln\left(  \frac{m_{g}^{2}}{4\mu_{0}^{2}}\right)  +\frac{1}{2}\right]
\end{equation}
or
\begin{equation}
\frac{\Lambda_{0}\left(  \mu_{0}\right)  }{8\pi G}=-\frac{m_{0}^{4}\left(
M\right)  }{128\pi^{2}}\left[  \ln\left(  \frac{m_{0}^{2}\left(  M\right)
}{4\mu_{0}^{2}}\right)  +\frac{1}{2}\right]  .
\end{equation}
Again we define a dimensionless variable%
\begin{equation}
x=\frac{m_{g}^{2}}{4\mu_{0}^{2}}\qquad\Longrightarrow\qquad\frac{\Lambda
_{0,0}\left(  \mu_{0},x\right)  }{8\pi G}=-\frac{G\mu_{0}^{4}}{\pi}%
x^{2}\left[  \ln\left(  x\right)  +\frac{1}{2}\right]  . \label{Lambdans0}%
\end{equation}
The formal expression of Eq.$\left(  \ref{Lambdans0}\right)  $ is very close
to Eq.$\left(  \ref{LambdansM}\right)  $ and indeed the extrema are in the
same position of the scale variable $x$, even if the meaning of the scale is
here different. $\Lambda_{0,0}\left(  \mu_{0},x\right)  $ vanishes for $x=0$
and $x=4\exp\left(  -\frac{1}{2}\right)  $. In this range, $\Lambda
_{0,0}\left(  \mu_{0},x\right)  \geq0$ and it has a minimum located in%
\begin{equation}
\bar{x}=\frac{1}{e}\qquad\Longrightarrow\qquad m_{g}^{2}=\frac{4\mu_{0}^{2}%
}{e} \label{min}%
\end{equation}
and%
\begin{equation}
\Lambda_{0,0}\left(  \mu_{0},\bar{x}\right)  =\frac{G\mu_{0}^{4}}{2\pi e^{2}}%
\end{equation}
or
\begin{equation}
\Lambda_{0,0}\left(  \mu_{0},\bar{x}\right)  =\frac{G}{32\pi}m_{g}^{4}%
=\frac{G}{32\pi}m_{0}^{4}\left(  M\right)  .
\end{equation}
Finally the case 3 ) leads to%
\begin{equation}
\frac{\Lambda_{0}\left(  \mu_{0},r\right)  }{8\pi G}\simeq-\frac{m_{0}%
^{4}\left(  M\right)  }{128\pi^{2}}\left[  \ln\left(  \frac{m_{0}^{2}\left(
M\right)  }{4\mu_{m}^{2}}\right)  +\frac{1}{2}\right]  ,
\end{equation}
where we have introduced another intermediate scale%
\begin{equation}
\mu_{m}^{2}=\mu_{0}^{2}\exp\left(  -\frac{3m_{g}^{4}}{2m_{0}^{4}\left(
M\right)  }\right)  .
\end{equation}
By repeating the same procedure of previous cases, we define%
\begin{equation}
x=\frac{m_{0}^{2}\left(  M\right)  }{4\mu_{m}^{2}}\qquad\Longrightarrow
\qquad\Lambda_{0,m}\left(  \mu_{0},x\right)  =-\frac{G\mu_{m}^{4}}{\pi}%
x^{2}\left[  \ln\left(  x\right)  +\frac{1}{2}\right]  . \label{Lambdansm}%
\end{equation}
Also this case has a maximum for%
\begin{equation}
\bar{x}=\frac{1}{e}\qquad\Longrightarrow\qquad m_{0}^{2}\left(  M\right)
=\frac{4\mu_{m}^{2}}{e}=\frac{4\mu_{0}^{2}}{e}\exp\left(  -\frac{3m_{g}^{4}%
}{2m_{0}^{4}\left(  M\right)  }\right)  .
\end{equation}
and%
\begin{equation}
\Lambda_{0,m}\left(  \mu_{0},\bar{x}\right)  =\frac{G\mu_{m}^{4}}{2\pi e^{2}%
}=\frac{G\mu_{0}^{4}}{2\pi e^{2}}\exp\left(  -\frac{3m_{g}^{4}}{m_{0}%
^{4}\left(  M\right)  }\right)
\end{equation}
or
\begin{equation}
\Lambda_{0,M}\left(  \mu_{0},\bar{x}\right)  =\frac{G}{32\pi}m_{0}^{4}\left(
M\right)  \exp\left(  \frac{3m_{g}^{4}}{m_{0}^{4}\left(  M\right)  }\right)  .
\end{equation}

\textbf{Remark }Note that in any case, the maximum of $\Lambda$ corresponds to
the minimum of the energy density.

A quite curious thing comes on the estimate on the \textquotedblleft square
graviton mass\textquotedblright, which in this context is closely related to
the cosmological constant. Indeed, from Eq.$\left(  \ref{min}\right)  $
applied on the square mass, we get%
\begin{equation}
m_{g}^{2}\propto\mu_{0}^{2}\simeq10^{32}GeV^{2}=10^{50}eV^{2},
\end{equation}
while the experimental upper bound is of the order%
\begin{equation}
\left(  m_{g}^{2}\right)  _{exp}\propto10^{-48}-10^{-58}eV^{2},
\end{equation}
which gives a difference of about $10^{98}-10^{108}$ orders. This discrepancy
strongly recall the difference of the cosmological constant estimated at the
Planck scale with that measured in the space where we live.

\end{document}